\documentclass[aps,twocolumn,pra,superscriptaddress]{revtex4}
\usepackage{epsfig,graphicx,times}
\usepackage{amstext}
\usepackage{amsmath}
\usepackage{lipsum}
\usepackage{amssymb}
\usepackage{float}
\usepackage{graphicx}
\usepackage{latexsym}
\usepackage{bm}
\usepackage{epstopdf}
\usepackage{appendix}
\usepackage[colorlinks,citecolor=blue, linkcolor=blue,hyperindex,CJKbookmarks]{hyperref}

\begin{document}

\title{Effect of Atom-Oscillator Interaction on the Aging Transition in Coupled Oscillators}
\author{Huining Zhang}
\affiliation{Center for Quantum Sciences and School of Physics, Northeast Normal University, Changchun 130024, China}
\author{X. Z. Hao}
\affiliation{Center for Quantum Sciences and School of Physics, Northeast Normal University, Changchun 130024, China}
\author{X. X. Yi\footnote{yixx@nenu.edu.cn}}
\affiliation{Center for Quantum Sciences and School of Physics, Northeast Normal University, Changchun 130024, China}
\affiliation{Center for Advanced Optoelectronic Functional Materials Research, and Key Laboratory for UV Light-Emitting Materials and Technology of Ministry of Education, Northeast Normal University, Changchun 130024, China}

\date{\today}

\begin{abstract}
Oscillators are often  employed as a model of radiation fields, which  may couple to an atom and play  an important role for creating and manipulating non-classical states in quantum metrology, quantum simulation and quantum information. Aging transitions in coupled oscillators have been studied extensively in both the classical and quantum contexts. It is well known that the onset of aging transitions can be modulated  by the dissipative coupling between oscillators. In this study, we propose an alternative way to modulate the aging transition through  coherent couplings between a  two-level atom and the oscillators. Our findings reveal that, compared to atom-free systems in both classical and quantum regimes, the atom-oscillator coherent interaction reduces the
inactive-to-total oscillator ratio required for aging transitions. Analytical results of the transition for both the classical oscillators and  quantum oscillators
suggest  that the decay rate of the atom and the atom-oscillator coupling strength jointly change the aging transition point. The  physics behind the observation is also elucidated in this manuscript. Our research introduces a readily implementable strategy for manipulating aging transitions in more intricate systems, thereby advancing the control and understanding of
these critical transitions in quantum technologies.
\end{abstract}
\maketitle

\section{INTRODUCTION}
Coupled oscillators serve as an excellent platform to unravel a plethora of nonlinear dynamical phenomena in many systems, including  transportation systems \cite{Ge2004}, power grids \cite{Rohden2012}, biological organisms \cite{Cama2003}, and the other physical systems \cite{Hopf1982}. The structure of the coupled oscillators and the strength of the interactions  promote the development of various collective behaviors, such as synchronization \cite{Piko2001,Xu2019}, clustering \cite{Xie2014,Kori2014}, solitary states \cite{Jaros2018,Teich2019,Sath2018,Sathi2019}, chimera patterns \cite{Kura1996,Kura1997,Zhu2014,Gopal2014}, and oscillation quenching \cite{Sax2012,Schn2015,Baner2018}. These facts imply  that the oscillators exhibiting complex dynamical behaviors may be coupled in diverse ways leading to many kinds  of collective behaviors.

Recently, a significant amount of research has focused on aging transitions \cite{Daido2004,Bandy2023,Zhang2024,Majhi2024,Daido2007,Daido2008,Sun2019,Sun2017,Sath2019,Sath2022,Ponrasu2020,Biswas2022,Sahoo2023,
Singh2020,Rahman2017,Rakshit2020,Zou2021}, and a distinct collective behavior was exhibited in coupled systems (such as oscillators or qubits), following the pioneering studies conducted by Daido \textit{et al} \cite{Daido2004}. Take the coupled oscillators as an example. In the aging progress, some oscillators lose their self-oscillatory nature and become inactive. When  examine what happens as the ratio of such inactive elements increases in an ensemble of globally coupled active and inactive oscillators, they found that as the ratio of these inactive elements (denoted by $p$) increases, the macroscopic oscillation gradually decreases and  ultimately undergoes  a transition to a stationary state when the ratio exceeds a critical value $p_c$. Later this phenomenon was extended to an ensemble of quantum oscillators  \cite{Bandy2023}. By treating oscillators as bosons, it was shown that quantum oscillator aging does not be characterized by a complete cessation into a collapsed state, but rather by a  significant reduction in the mean boson number. There exists a critical ``knee'' value of the fraction of the inactive oscillators around which the aging of coupled quantum oscillators exhibits two distinct trends of decrement in the mean boson number. Very recently, the aging of coupled qubits networks has also been explored \cite{Zhang2024}. The aging transition manifests as a sudden drop in the average excited-state population when the ratio of inactive qubits exceeds a critical value \cite{Zhang2024}.

We observed  that in exploring the aging transition of coupled-oscillator systems, all publications  have thus far concentrated solely on analyzing the effects of inter-oscillator interactions on aging transitions \cite{Daido2004,Bandy2023,Majhi2024,Daido2007,Daido2008,Sun2019,Sun2017,Sath2019,Sath2022,Ponrasu2020,Biswas2022,Sahoo2023,
Singh2020,Rahman2017,Rakshit2020,Zou2021}. For a better understanding of the aging in  complex systems, one needs to further consider  couplings  between different types of systems, for instance, atom-oscillator couplings. The coupling between atoms (or quantum emitters) and oscillators (e.g., mechanical resonators, electromagnetic field modes, or collective excitations) lies at the heart of hybrid quantum systems. Atom-oscillator couplings are ubiquitous in modern physical systems, such as in circuit optomechanics \cite{Hei2014,Pirk2015,Liao2014}, nitrogen-vancancy (NV) center-mechanical oscillator hybrid systems \cite{Li2020}, and the implementation of entangling operations between atomic and photonic qubits, as well as between pairs of photonic qubits \cite{Duan2004,Duan2005}. Additionally, the superradiant phase transition in the atom-boson systems have attracted a wide range of attention \cite{Hwang2015} for decades. Recognizing the significant role of atom-oscillator coupling, we are interested  in its effect on the aging transition of coupled oscillators, and want to gain deeper insights into the aging behavior of complex systems.

The system under consideration  consists of $N$ all-to-all coupled active-inactive oscillators and  a two-level atom. Each oscillator is coherently coupled to the atom. We will study the effect of atom-oscillator couplings on the aging transition of the oscillators in both the classical and quantum regimes. In the classical regime, with respect  to the scenario without atom-oscillator coupling, the aging transition can occur at a lower inactive-total oscillators ratio. The minimum value of $p$ that allows the system to undergo an aging transition can be analytically derived from the linear stability analysis. The similar effects induced by the atom-oscillator coupling  also manifest in the quantum regime. Our numerical simulations have revealed that the couplings between the atom and quantum oscillators can also lead to an aging transition  at a smaller value of $p$ with respect  to the case without couplings. This phenomena can be  explained by adiabatically eliminating the atomic degrees of freedom from the system dynamics. Moreover, we demonstrate how the atom-oscillator couplings and the decay rate of the atom modulate the aging transition.

This work is organized as follows. In sec.~\ref{sec1}, we introduce our theoretical model. The main results of this work are presented in Sec.~\ref{sec2} for classical regime and in Sec.~\ref{sec3} for quantum regime. In these two sections, we also reveal the physics behind the observation that, compared to the system without an atom, the aging transition occurs at a lower ratio of inactive oscillators. Finally, we conclude our results  in Sec.~\ref{sec4}. Appendix~\ref{appendixA} demonstrates the reasonableness to treat the oscillator as  classical when the nonlinear damping rate is very small. Appendix~\ref{appendixB} gives the reason for the same decay rate of the normalized order parameter $Q_c$ for different values of the atom-oscillator coupling strength. Details for the derivation of the Fock distribution of the quantum active and inactive oscillators after the system undergoing the aging transition are given in Appendix~\ref{appendixC}.

\section{MODEL}\label{sec1}

For $N$ quantum oscillators interacting with a two-level atom, the Hamiltonian for the system in the rotating wave approximations is given by (with $\hbar=1$)
\begin{equation}
H=\frac{1}{2}\omega_0\sigma_z+\sum_{n=1}^N\omega_n a_{n}^{\dag}a_n+g\sum_{n=1}^N(a_{n}^{\dag}\sigma_{-}+a_{n}\sigma_{+}),
\label{Hamiltonian1}
\end{equation}
$\omega_n$ and $\omega_0$ represent the  energy  of the $n$th oscillator and the atom, respectively. $a_{n}^{\dag}$ and $a_{n}$ are the creation and annihilation operators for the $n$th oscillator. $\sigma_{z} = |e\rangle\langle e| - |g\rangle\langle g|$, $\sigma_{+} = |e\rangle\langle g|$, and $\sigma_{-} = |g\rangle\langle e|$ are Pauli operators, where $|e\rangle$ and $|g\rangle$ denote the excited and ground states of the atom. The first and second terms of Eq.~(\ref{Hamiltonian1}) correspond to the free Hamiltonian of the two-level atom and quantum oscillators, respectively. The third term describes the interaction Hamiltonian between the atom and the oscillators, with coupling strength $g$.

In order to address the problem easily, we transform the system into the interaction picture by defining $U=\text{exp}[-i(\frac{1}{2}\omega_0\sigma_z+\sum_{n=1}^N\omega_n a_{n}^{\dag}a_n)t]$, the Hamiltonian is then
\begin{equation}
\begin{aligned}
H^{'}&=i\frac{dU^\dag}{dt}U+U^\dag HU\\
&=g\sum_{n=1}^N(a_{n}^{\dag}\sigma_{-}e^{-i\Delta t}+a_{n}\sigma_{+}e^{i\Delta t}),
\label{Hamiltonian2}
\end{aligned}
\end{equation}
where $\Delta=\omega_0-\omega_n$. Here we consider that each oscillator has the same energy and resonantly  couples to the atom, i.e., $\Delta=0$.

Quantum master equation of $N$ coupled quantum oscillators and a two-level atom under diffusive global coupling is given by \cite{Bandy2023,Lee2014,Ishi2017}
\begin{equation}
\begin{aligned}
\dot{\rho}&=-i[H^{'},\rho]+\sum_{n=1}^{N_a} \gamma_{n} \mathcal{D}[a_{n}^{\dag}](\rho) + \sum_{n=N_a+1}^{N} \gamma_{n} \mathcal{D}[a_{n}](\rho)\\&+\sum_{n=1}^{N}\kappa\mathcal{D}[a_{n}^2](\rho)+
{\sum_{m,n}}^{'}\frac{V}{N}\mathcal{D}[a_{m}-a_{n}](\rho)+J\mathcal{D}[\sigma_{-}](\rho),
\label{master equation}
\end{aligned}
\end{equation}
where $\rho$ denotes the system's density matrix, and $\mathcal{D}[\hat{\mathcal O}](\rho) = \hat{\mathcal O}\rho\hat{\mathcal O}^{\dagger} - \frac{1}{2}\{\hat{\mathcal O}^{\dagger}\hat{\mathcal O},\rho\}$. The second and third terms $\mathcal{D}[a_{n}^{\dag}](\rho)$ and $\mathcal{D}[a_{n}](\rho)$ in the quantum master equation distinguish between active and inactive oscillators. Active oscillators undergo incoherent transitions from the lower energy level to the higher energy level with rates $\gamma_n, n=1,2,3,...,N_a$. Conversely, inactive oscillators transition from the higher energy level to the lower energy level with rates $\gamma_n, n=N_a+1, ... , N$. These notations are the same as those used for quantum oscillators \cite{Bandy2023}. $\kappa$ corresponds to the nonlinear damping. The term involving $V$ denotes the dissipative coupling between two oscillators. The sum ${\sum_{m,n}}^{'}$ indicates a summation over distinct oscillators. The coefficient $J$ is the decay rate of the atom. With this notation, \(N\) coupled oscillators are partitioned into two groups: \(N_a\) active elements and \(N_i = N - N_a\) inactive elements, characterizing by  the ratio of inactive to the total oscillators defined by \(p = \frac{N_i}{N} = \frac{N - N_a}{N}\).  For convenience, we set the group of active elements to $n\in\{1,...,N(1-p)\}\equiv N_a$ and that of inactive elements to $n\in\{N(1-p)+1,...,N\}\equiv N_i$.

\section{Aging transition in classical regime}\label{sec2}
\begin{figure}[t]
	\centering
	\includegraphics[width=0.47\textwidth]{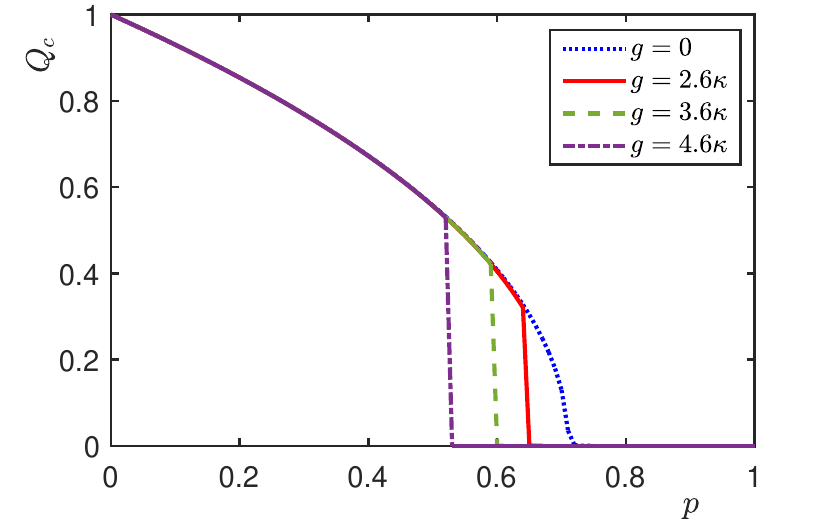}
	\caption {The normalized order parameter $Q_c$ as a function of the ratio of the inactive oscillators to the total oscillators $p$ for different values of atom-oscillator coupling strength $g$. For active oscillators $\gamma_n=a=80\kappa$ and for inactive oscillators $\gamma_n=b=40\kappa$. The other parameters are $N=100, J=300\kappa, V=300\kappa$.}
	\label{fig1}
\end{figure}

In this section, we present the results for the oscillators in the classical regime. When $\kappa$ is small with respect  to $\gamma_n$ (i.e., $\kappa\ll\gamma_n$), the oscillators can be considered classical and one can approximate $a_n\equiv\alpha_n$. We will show  that the results obtained in the classical regime are in good agreement  with that in the quantum regime under specific conditions. For details, see Appendix \ref{appendixA}. Starting from the master equation (\ref{master equation}), the time evolution of the operator's expectation values for the atom-oscillator system can be obtained by  $\dot{\langle\mathcal O\rangle}=\text{Tr}(\dot{\rho}\mathcal O),$
\begin{equation}
\begin{aligned}
\dot{\langle\sigma_{-}\rangle}&=-ig\sum_{n=1}^{N}(\langle a_n\rangle-2\langle a_n\sigma_{+}\sigma_{-}\rangle)-\frac{J}{2}\langle\sigma_{-}\rangle,\\
\dot{\langle\sigma_{+}\sigma_{-}\rangle}&=-ig\sum_{n=1}^{N}(\langle a_n\sigma_{+}\rangle-\langle a_{n}^{\dag}\sigma_{-}\rangle)-J\langle\sigma_{+}\sigma_{-}\rangle,\\
\dot{\langle a_n\rangle}&=-ig\langle\sigma_{-}\rangle\pm\frac{\gamma_n}{2}\langle a_n\rangle-\kappa\langle a_{n}^\dag a_{n}^2\rangle\\&+\frac{V}{N}\sum_{m=1}^N(\langle a_m\rangle-\langle a_n\rangle),
\label{equation1}
\end{aligned}
\end{equation}
the sign in front of $\frac{\gamma_n}{2}$ is plus for active and minus for inactive oscillators. For weak atom-oscillator couplings,  we may take the following  mean-field approximations  \cite{Kubo1962}
\begin{equation}
\begin{aligned}
\langle a_n\sigma_{+}\sigma_{-}\rangle&\approx\langle a_n\rangle\langle\sigma_{+}\sigma_{-}\rangle,\\
\langle a_n\sigma_{+}\rangle&\approx\langle a_n\rangle\langle\sigma_{+}\rangle,
\label{equation2}
\end{aligned}
\end{equation}
In the classical regime, we can replace the expectation $\langle a_n\rangle$ with a complex amplitude $\alpha_n$, as well as $\langle a_{n}^\dag a_{n}^2\rangle\rightarrow|\alpha_n|^2\alpha_n$ in the equation of motion for $\langle a_n\rangle$ given by  Eq.~(\ref{equation1}). Assuming that in each group, all oscillators are in the same state and setting $\alpha_n=A$ and $\gamma_n=a$ for all active oscillators as well as $\alpha_n=I$ and $\gamma_n=b$  for all inactive oscillators, we obtain
\begin{equation}
\begin{aligned}
\dot{\langle\sigma_{-}\rangle}&=-ig\sum_{n=1}^{N}\alpha_n
(1-2\langle\sigma_{+}\sigma_{-}\rangle)-\frac{J}{2}\langle\sigma_{-}\rangle,\\
\dot{\langle\sigma_{+}\sigma_{-}\rangle}&=-ig\sum_{n=1}^{N}(\alpha_n\langle\sigma_{+}\rangle-\alpha_{n}^{*}
\langle\sigma_{-}\rangle)-J\langle\sigma_{+}\sigma_{-}\rangle,\\
\dot{A}&=-ig\langle\sigma_{-}\rangle+(\frac{a}{2}-Vp-\kappa|A|^2)A+VpI,\\
\dot{I}&=-ig\langle\sigma_{-}\rangle+(-\frac{b}{2}-Vq-\kappa|I|^2)I+VqA,
\label{equation3}
\end{aligned}
\end{equation}
where $q\equiv1-p$. When there is no coupling in the system (namely, $V=0,g=0$), the active oscillator exhibits a stable limit-cycle oscillation in the Schr$\ddot{\text o}$dinger picture, whereas the inactive oscillator is non-self-oscillatory. In the classical regime, we explore the aging transition through the  normalized order parameter $Q_c=|R(p)|/|R(0)|$, where $R\equiv N^{-1}\sum_{n=1}^{N}\alpha_n$ and show the results in Fig.~\ref{fig1}.

Fig.~\ref{fig1} shows the order parameter as a function of $p$ for some values of $g$. It is clear that $Q_c$ decreases
monotonically with $p$ and transitions from a finite value to zero, indicating the transition from an oscillatory state to an aging state when $p$ exceeds the critical ratio $p_c$. For example, $p_c=0.64$ when $g=2.6\kappa$. Here the aging state corresponds to the state where the oscillators' complex amplitudes $A=I=0$. Without atom-oscillator coupling ($g=0$), the system reduces to the results given by Daido and Nakanishi \cite{Daido2004}, leading  to the aging transition in the globally coupled classical oscillators system. With atom-oscillator coupling and  the parameters chosen, the aging transition occurs at a critical value $p_c$ smaller  than the case without atom-oscillator couplings. Moreover, the critical ratio $p_c$ decreases with the coupling strength $g$ increases.
\begin{figure}[t]
	\centering
	\includegraphics[width=0.47\textwidth]{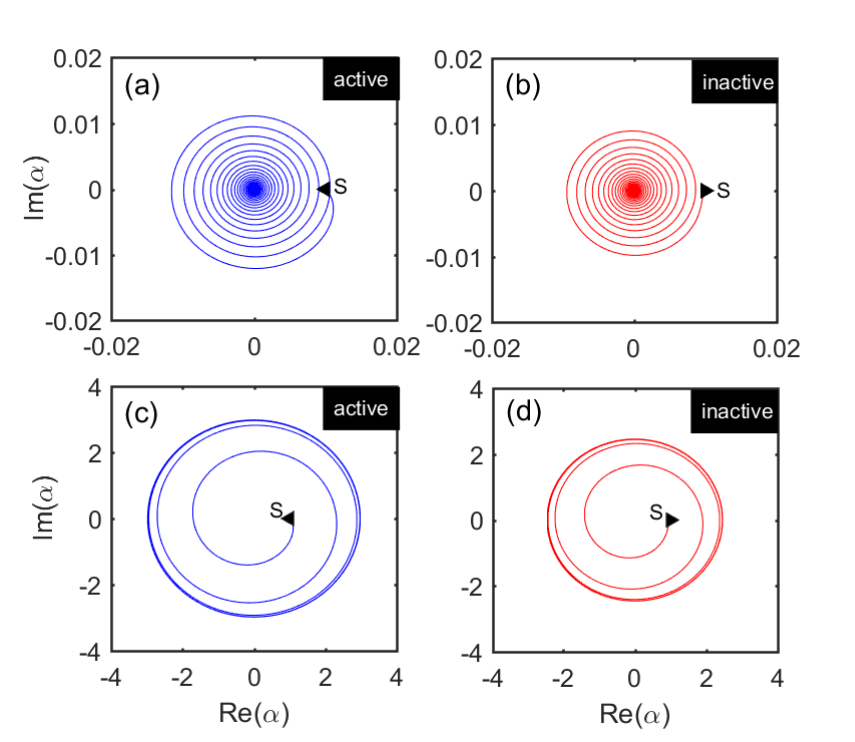}
	\caption {Time evolution of the complex amplitude $\alpha$ for the active and inactive oscillators in the phase space in the Schr$\ddot{\text o}$dinger picture under the initial condition  (a)-(b) $A_0=I_0=0.01$ (marked by S). (c)-(d) $A_0=I_0=1$. The atom is initially in the excited state. $p=0.59, g=3.6\kappa$. Other parameters are the same as in Fig.~\ref{fig1}.}
	\label{fig2}
\end{figure}

To reveal  the physics  behind these observations, we numerically study the behaviors of the active and the inactive oscillators  under different initial conditions. Specifically, for the initial condition of complex amplitudes $A_0=I_0=0.01$, the oscillators ultimately converge to the fixed point $A=I=0$ of Eq.~(\ref{equation3}) [see Fig.~\ref{fig2}(a)-(b)]. However, when $A_0=I_0=1$, the oscillators eventually exhibit periodic oscillations [see Fig.~\ref{fig2}(c)-(d)]. This means that the dynamics of the system are crucially dependent on the  initial conditions, which in turn determines whether the system evolves into the aging state at a particular value $p_c$ of $p$. This motivated us to explore the threshold of $p$ that stabilizes the fixed point $A=I=0$.  We should note that  reaching this threshold for $p$  does not necessarily mean the system  immediately transit into an aging state, as this depends sharply  upon the initial conditions. Different initial conditions may lead to different critical points $p_c$. Nevertheless, we can define  a threshold that represents the minimum value of $p_c$ (will be denoted by $p_\text{cmin}$) at which the system is capable of undergoing an aging transition.

\begin{figure}[t]
	\centering
	\includegraphics[width=0.47\textwidth]{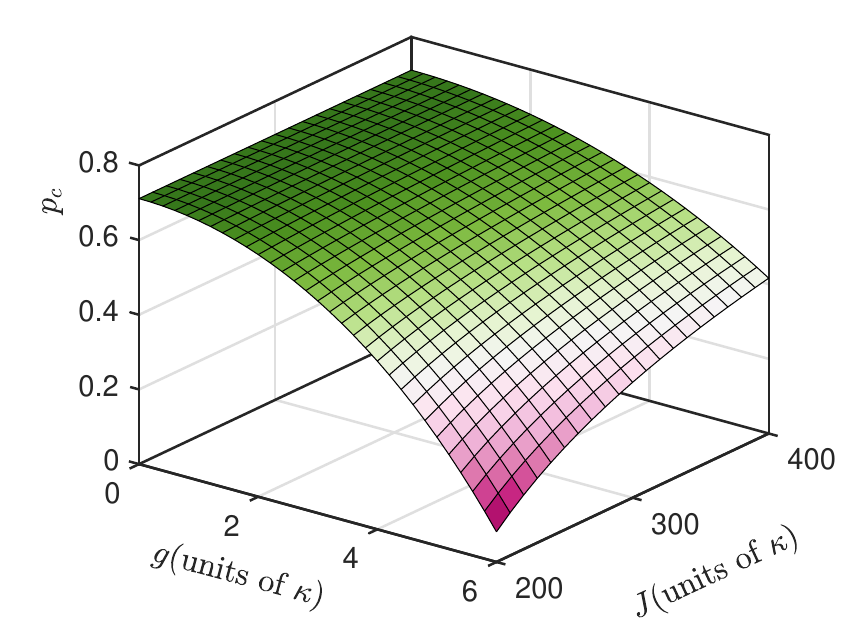}
	\caption {The dependence of the threshold $p_\text{cmin}$ on the atom-oscillator coupling strength $g$ and decay rate of the atom $J$. Other parameters chosen  are the same as in Fig.~\ref{fig1}.}
	\label{fig3}
\end{figure}

The threshold $p_\text{cmin}$ can be analytically determined from the linear stability analysis \cite{Strogatz1994}. To find the threshold, we discuss the stability of the aging state. It follows that the aging transition could occur when the trivial fixed point $\langle\sigma_{-}\rangle=\langle\sigma_{+}\sigma_{-}\rangle=A=I=0$ of Eq.~(\ref{equation3}) is stabilized as $p$ is increased from zero. Let $\xi_1$ be a deviation  from the fixed point $\langle\sigma_{-}\rangle=0$, $\xi_2$  a deviation from the fixed point $\langle\sigma_{+}\sigma_{-}\rangle=0$, $\xi_3$ a deviation from the fixed point $A=0$, and $\xi_4$ be a deviation from the fixed point $I=0$. The equation for these deviations  can be  derived from Eq.~(\ref{equation3})  as follows
\begin{equation}
\begin{aligned}
\dot{\xi_1}&=-J/2\xi_1-ig[N(1-p)\xi_3+Np\xi_4],\\
\dot{\xi_2}&=-J\xi_2,\\
\dot{\xi_3}&=-ig\xi_1+(\frac{a}{2}-Vp)\xi_3+Vp\xi_4,\\
\dot{\xi_4}&=-ig\xi_1+Vq\xi_3+(-\frac{b}{2}-Vq)\xi_4,
\label{equation4}
\end{aligned}
\end{equation}
which can be written in  a more compact form
\begin{equation}
\begin{pmatrix}
\dot{\xi_1} \\
\dot{\xi_2} \\
\dot{\xi_3} \\
\dot{\xi_4}
\end{pmatrix}
=M
\begin{pmatrix}
\xi_1 \\
\xi_2 \\
\xi_3 \\
\xi_4
\end{pmatrix}
\end{equation}
with $M$ defined by,
\begin{equation}
M=
\begin{pmatrix}
-\frac{J}{2} & 0 & -igN(1-p) & -igNp \\
0 & -J & 0 & 0 \\
-ig & 0 & \frac{a}{2}-Vp & Vp \\
-ig & 0 & V(1-p) & -\frac{b}{2}-V(1-p)
\end{pmatrix}.
\end{equation}
By applying the ansatz $\xi_j(t)\propto e^{\lambda t}$ into Eq.~(\ref{equation4}), the characteristic equation governing the stability of the fixed points is given by $|M-\lambda I|=0$, i.e.,
\begin{equation}
\begin{aligned}
(-J-\lambda)(\lambda^3+c_2\lambda^2+c_1\lambda+c_0)=0,
\label{equation5}
\end{aligned}
\end{equation}
where the coefficients $c_i$ read,
\begin{equation}
\begin{aligned}
c_0&=\frac{1}{8}[(-4bg^2N(-1+p)+8g^2NV+2bJpV\\&-a(bJ+4g^2Np-2J(-1+p)V))],\\
c_1&=\frac{1}{4}[bJ+4g^2N+2(J+bp)V\\&-a(b+J+2V-2pV)],\\
c_2&=\frac{1}{2}(-a+b+J)+V,
\end{aligned}
\end{equation}
we can deduce the threshold using the coefficients of the characteristic equation [Eq.~(\ref{equation5})] following the Routh-Hurwitz stability criterion \cite{Liu1994}. In our manuscript, all parameters are non-negative real numbers. With the parameters range chosen $(2g^2N<JV)$, we arrive at the minimum value of $p_c$ that allows the aging transition to occur
\begin{equation}
\begin{aligned}
p_\text{cmin}=\frac{(4g^2N-aJ)(b+2V)}{2(2g^2N-JV)(a+b)}.
\label{equation6}
\end{aligned}
\end{equation}
When $p$ goes beyond this threshold, the oscillators collapse into global non self-oscillatory state by carefully choosing initial conditions. In the limiting case where $g$ approaches 0, the threshold $p_\text{cmin}$ converges to $\frac{a(b+2V)}{2V(a+b)}$, which is the same as the critical ratio for a diffusively coupled global network \cite{Daido2004}. Additionally, we find that the normalized order parameter $Q_c$ exhibits the same decay rate for different values of the atom-oscillator coupling strength $g$ before the aging transition occurs as shown in Fig.~\ref{fig1}. This phenomenon can be understood from the details provided in Appendix \ref{appendixB}. The dependence  of $p_\text{cmin}$ on the atom-oscillator coupling strength $g$ and the decay rate of the atom $J$ is illustrated in Fig.~\ref{fig3}. For a fixed decay rate $J$, the threshold decreases monotonically   with  $g$ increasing. Consequently, the aging transition can occur earlier (i.e., for smaller $p$) with respect to the case of $g=0$, leading to a decrease in the robustness of the system due to the introduction of the interactions between the atom and the oscillators.  However, for a fixed coupling strength $g$, increasing the decay rate of the atom can enhance the robustness of the system.

We observe that, depending on parameter values, atom-oscillator interactions might cause sudden suppression of macroscopic oscillations in the coupled-oscillator system. This abrupt  transition is a significant phenomenon, particularly in the context of practical systems.  This transition is distinguished by its catastrophic nature, wherein the magnitude of the order parameter does not afford any prior indication of the imminent collapse of the system. Hence, predicting the lower bound ($p_\text{cmin}$) of the critical ratio for the occurrence of the aging transition is of paramount importance.

\section{Aging transition in quantum regime}\label{sec3}
We are interested in what happens to the aging behavior of the above models in the quantum oscillators as well. In the quantum regime, the nonlinear damping rate is large with respect to the linear pumping rate. The oscillator is approximately occupied near the ground state and hence its  quantum properties can not be ignored. Therefore, the annihilation and creation operators can no longer be simply replaced by a number. To study this, we go back to Eq.~(\ref{master equation}). For a large number of particles, one can factorize the density matrix of the many-body system as $\rho\approx\rho_a\bigotimes\rho_1\bigotimes\rho_2\bigotimes\cdots\bigotimes\rho_n$ \cite{Lee2014,Ishi2017}. $\rho_a$ and $\rho_n (n=1,...,N)$ denote the one-body density matrix of the atom and quantum oscillators, respectively. In the following, $a_n$ is represented by $a$. Since all oscillators
within each group are identical, their time evolution can be described in terms of two oscillators  with density matrices $\rho_A$ (active) and $\rho_I$ (inactive) coupled to the atom and the mean-field $M_A$ and $M_I$
\begin{equation}
\begin{aligned}
\dot{\rho_a}&=-i[g(M\sigma_{+}+M^*\sigma_{-}),\rho_a]+J\mathcal{D}[\sigma_{-}](\rho_a),\\
\dot{\rho_A}&=-i[g(\langle\sigma_{+}\rangle a+a^{\dag}\langle\sigma_{-}\rangle),\rho_A]+\gamma_{n} \mathcal{D}[a^{\dag}](\rho_A)\\&+\kappa\mathcal{D}[a^2](\rho_A)+\frac{2V(N-1)}{N}\mathcal{D}[a](\rho_A)
\\&+\frac{V}{N}(M_A[a^\dag,\rho_A]-{M_A}^*[a,\rho_A]),\\
\dot{\rho_I}&=-i[g(\langle\sigma_{+}\rangle a+a^{\dag}\langle\sigma_{-}\rangle),\rho_I]+\gamma_{n} \mathcal{D}[a](\rho_I)\\&+\kappa\mathcal{D}[a^2](\rho_I)+\frac{2V(N-1)}{N}\mathcal{D}[a](\rho_I)
\\&+\frac{V}{N}(M_I[a^\dag,\rho_I]-{M_I}^*[a,\rho_I]),
\label{equation7}
\end{aligned}
\end{equation}
$\dot{\rho_A} (\dot{\rho_I})$ represents the reduced master equation for active (inactive) oscillators. The mean-fields  are given by
\begin{equation}
\begin{aligned}
M&={\sum_{m}}{\langle a\rangle}_m=
N(1-p)\text{Tr}(a\rho_A)+Np\text{Tr}(a\rho_I),\\
M_A&={\sum_{m}}^{'}{\langle a\rangle}_m=
[N(1-p)-1]\text{Tr}(a\rho_A)+Np\text{Tr}(a\rho_I),\\
M_I&={\sum_{m}}^{'}{\langle a\rangle}_m=
N(1-p)\text{Tr}(a\rho_A)+(Np-1)\text{Tr}(a\rho_I),
\label{equation8}
\end{aligned}
\end{equation}
Here ${\sum_{m}}^{'}$ indicates that the sum does not include the condition $m=n$. For active (inactive) oscillators, the mean field $M_A (M_I)$ has different expressions as can be seen from Eq.~(\ref{equation8}). Mean-field theory yields a set of three closed nonlinear equations described by Eq.~(\ref{equation7}). Throughout this manuscript, we suppose  that the quantum oscillators describe bosons, then we can truncate the Hilbert space to a certain number in our numerical simulations. In the quantum regime, we consider the case: $\kappa>\gamma_n$, the higher  damping makes the oscillators to populate only a few excited state near the ground state \cite{Lee2013}. Thus, we truncate the Hilbert space to the boson number $n_0=4$, since other higher Fock states   quickly decay due to the dampings.
\begin{figure}[t]
	\centering
	\includegraphics[width=0.47\textwidth]{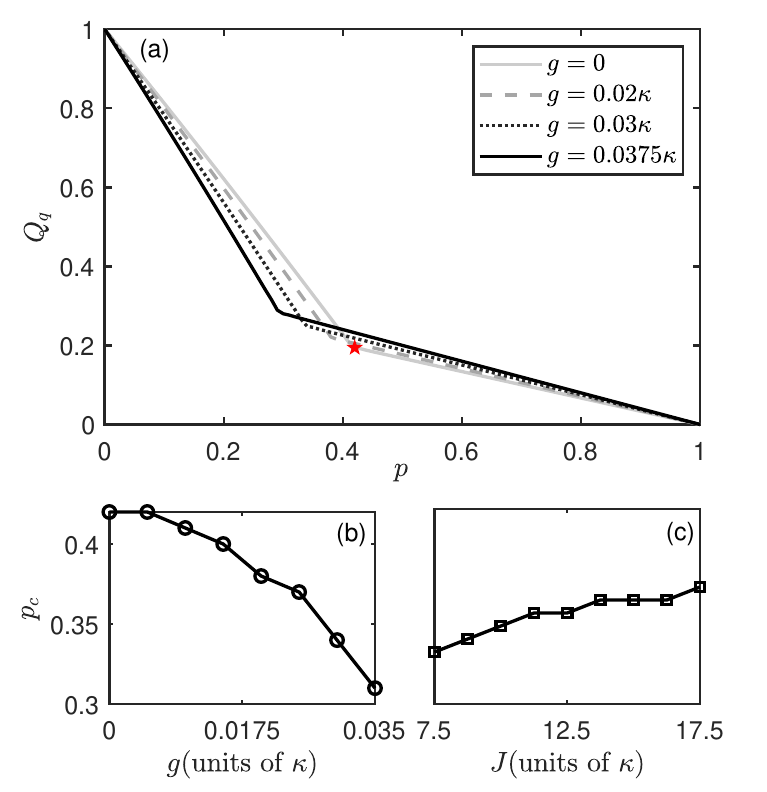}
	\caption {(a) Quantum aging transition: Dependence  of order parameter $Q_q$ on  $p$ for different atom-oscillator coupling strength $g$. (b) The dependence  of $p_c$ on atom-oscillator coupling strengths $g$ with $J=7.5\kappa$. (c) The dependence  of $p_c$ on atom decay rates $J$ with $g=0.03\kappa$. For active oscillators $\gamma_n=a=0.5\kappa$ and for inactive oscillators $\gamma_n=b=0.375\kappa$. The other parameters chosen are $N=100, J=7.5\kappa, V=3.75\kappa$.}
	\label{fig4}
\end{figure}

In the quantum regime, we study the aging of quantum oscillators by calculating the mean
boson number, defined by $\bar{n}_0=\frac{1}{N}\sum_n{\langle a^{\dag}a\rangle}_n$ \cite{Bandy2023}. To this end, we introduce the normalized mean boson number $Q_q=\bar{n}_0(p)/\bar{n}_0(0)$ as an order parameter. To explore whether the aging transition in quantum regime, upon interacting with atoms, displays analogous effects to those observed in classical regime, we analyze the behavior of the order parameter $Q_q$ as a function of $p$ for different  $g$, we plot the results  in Fig.~\ref{fig4}(a). Firstly, we observe that the curves exhibit two different decline trends on both sides of a critical ratio $p_c$ (marked by a star). The aging transition occurs when the inactive ratio increases to this critical point, indicating the beginning of the system to the aging state.  We identify  these critical points as aging transition points.  Within the chosen parameter range, we observe that, in contrast  to cases of $g=0$, the aging transition occurs at a smaller $p_c$ when the coupling constant $g$ is not zero. This behavior in agreement  with the characteristics observed in the classical regime.

We provide a clear explanation for this observation. Because the number of oscillators $N$ is large, it ensures the approximations $M\approx M_A\approx M_I$. In our manuscript, we assume that the decay rate of the atom $J$ is much larger than all other relevant parameters. Hence, we can eliminate the atomic degrees of freedom adiabatically \cite{Lugiato2015} by setting $\dot{\langle\sigma_{-}\rangle}=\dot{\langle\sigma_{+}\sigma_{-}\rangle}=0$ [$\dot{\langle \sigma_{-}\rangle}$=Tr($\dot{\rho_a}\sigma_{-}$)], which gives
\begin{equation}
\begin{aligned}
&\langle\sigma_{-}\rangle=\frac{-2gi}{J}M(1-2\langle\sigma_{+}\sigma_{-}\rangle),\\
&\langle\sigma_{+}\sigma_{-}\rangle=\frac{1}{2+\frac{J^2}{4g^2|M|^2}},
\label{reply_eq6}
\end{aligned}
\end{equation}
The decay rate of the atom $J$ is significantly larger than the atom-oscillator coupling strength $g$, such that $J^2\gg4g^2|M|^2$, and the population of the atom in the excited state tends to zero. Consequently, the expectation value of the atomic lowering operator can be approximated as $\langle\sigma_{-}\rangle\approx\frac{-2gi}{J}M$. By eliminating the atomic degrees of freedom from the system dynamics, we derive the master equation that governs the evolution of the oscillators
\begin{equation}
\begin{aligned}
\dot{\rho_A}&=\gamma_{n}\mathcal{D}[a^{\dag}](\rho_A)+\kappa\mathcal{D}[a^2](\rho_A)+\frac{2V(N-1)}{N}\mathcal{D}[a](\rho_A)
\\&+(\frac{V}{N}+G)(M_A[a^\dag,\rho_A]-{M_A}^*[a,\rho_A]),\\
\dot{\rho_I}&=\gamma_{n} \mathcal{D}[a](\rho_I)+\kappa\mathcal{D}[a^2](\rho_I)+\frac{2V(N-1)}{N}\mathcal{D}[a](\rho_I)
\\&+(\frac{V}{N}+G)(M_I[a^\dag,\rho_I]-{M_I}^*[a,\rho_I]),
\label{reply_eq7}
\end{aligned}
\end{equation}
where $G=-\frac{2g^2}{J}$. From Eq.~(\ref{reply_eq7}), we find that the  atom  modulates the coupling strength of each oscillator to  $M_A$ or $M_I$.  Specifically, $G<0$ results in a reduced effective coupling strength $V$ compared to the case where $g=0$. We verify that the steady-state results obtained by using Eq.~(\ref{reply_eq7}) are in good agreement with that based on Eq.~(\ref{equation7}). As shown in Fig.~\ref{fig4}(a), a negative $G$ shifts the critical ratio $p_c$ to the left  (i.e., decreases $p_c$). Further analysis of Fig.~\ref{fig4}(b) and (c) reveals that $p_c$ decreases as the increase of the atom-oscillator coupling strength. This occurs because a larger $g$ leads to a smaller effective coupling term $\frac{V}{N}+G$, thereby further decreasing the critical ratio $p_c$. Conversely, increasing the atomic decay rate $J$ enlarges $\frac{V}{N}+G$, causing  a shift in $p_c$ and making it more close  to the value when there is no atom-oscillator coupling. When $J$ is sufficiently large or $g$ is sufficiently small, the system's behavior aligns with the scenario where no atom-oscillator coupling exists.
\begin{figure}[t]
	\centering
	\includegraphics[width=0.5\textwidth]{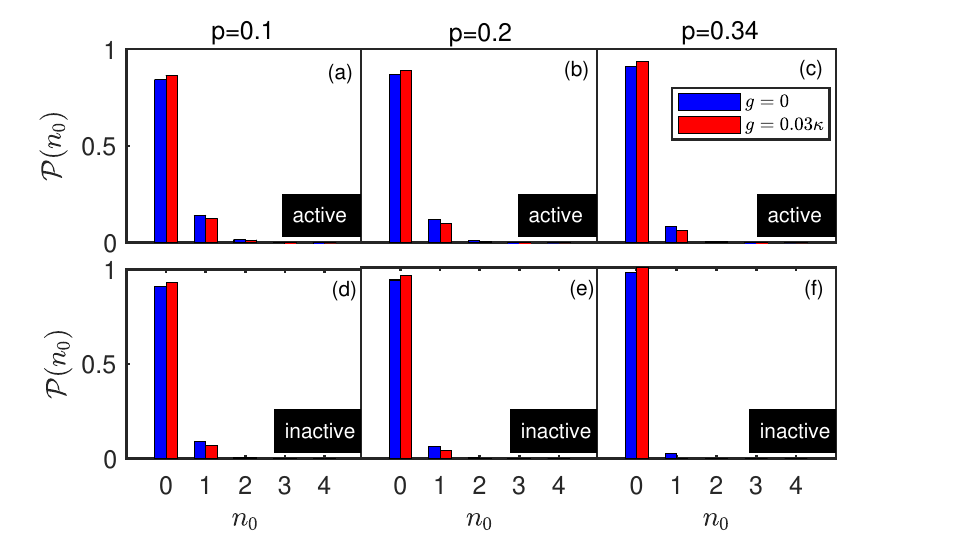}
	\caption {Population of  boson numbers $n_0$, $\mathcal P(n_0)$ for active (upper row) and inactive (lower row) oscillators  for different inactive-active oscillator ratio $p$ in the absence and presence of the atom-oscillator interaction. [(a), (d)] $p = 0.1$, [(b), (e)] $p = 0.2$. [(c), (f)] $p = 0.34$. Other parameters chosen are the same as in Fig.~\ref{fig4}.}
	\label{fig5}
\end{figure}

The scenario of quantum aging in the context of atom-oscillator interaction can be understood by closely examining the distribution of the system on Fock states. Fig.~\ref{fig5}(a)-(f) depict the distribution of Fock states for both active and inactive groups at different $p$. Compared to the uncoupled case ($g=0$), at the same value of $p$, we observe a remarkable feature: In the presence of coupling between the atom and the quantum oscillators, both active and inactive elements exhibit higher occupation numbers in the ground state. This suggests that the interaction between the atom and the oscillators moves  the probability distribution of the oscillators towards the ground state. As $p$ increases, the probability distribution continues to move toward the ground state. Consequently, when $p$ reaches a value that is lower than the critical point observed in the uncoupled case, the inactive oscillators fully occupy the ground state [see Fig.~\ref{fig5}(f)]. This marks  the occurrence of the aging transition.

Collecting aforementioned observations, we claimed that after the aging transition the inactive oscillators completely collapse to the ground state, whereas active oscillators do not fully relax to the ground state. Because for active components, there are two competitive channels: a single-boson gain and a double-boson loss. The presence of the single-boson gain process prevents the system from fully relaxing to the ground state, as it has the potential to excite the oscillator from a lower energy level to a higher one. In contrast, for inactive oscillators, both the two channels are dissipative, which allow for complete relaxation to the ground state. Moreover, beyond the critical point $p_c$, we find that the density matrix $\rho_A$ and $\rho_I$ have no off-diagonal elements,  this implies the lack of coherence among the quantum oscillators. A more detailed analysis is provided in Appendix \ref{appendixC}, where the analytical expressions for the Fock distribution of the active and inactive oscillators can be found.

\begin{figure}[t]
	\centering
	\includegraphics[width=0.5\textwidth]{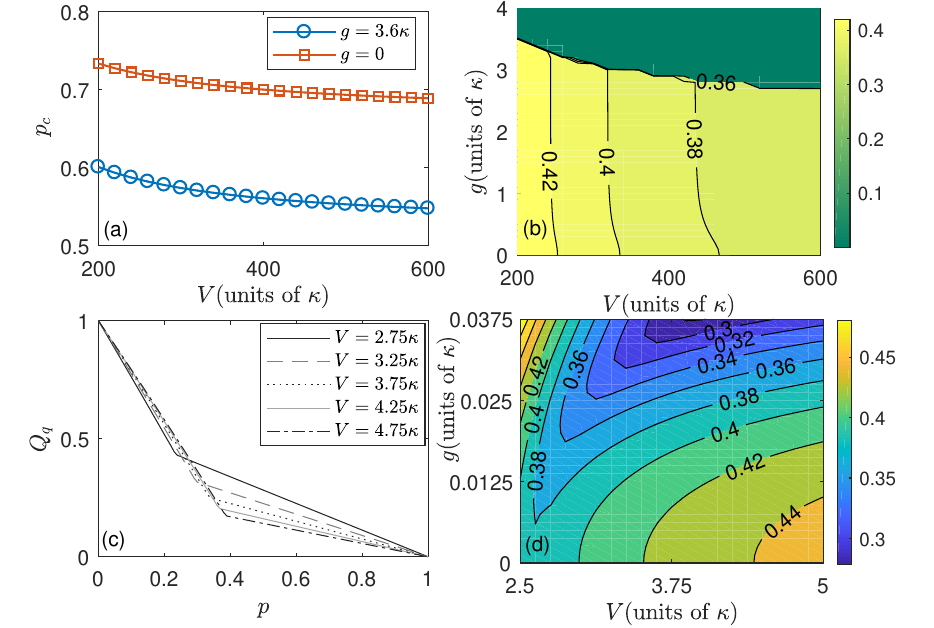}
	\caption {In the classical regime: (a) The dependence of the threshold $p_\text{cmin}$ on the dissipative coupling strength $V$ in the absence and presence of the atom-oscillator interaction. (b) Contour plot of the normalized order parameter $Q_c$ over the parameter space $g-V$. For active oscillators $\gamma_n=a=80\kappa$ and for inactive oscillators $\gamma_n=b=40\kappa$. The other parameters are $N=100, J=300\kappa$. In the quantum regime: (c) Dependence of the order parameter $Q_q$ on $p$ for different dissipative coupling strength $V$. (d) Contour plot of the normalized order parameter $Q_q$ over the parameter space $g-V$. For active oscillators $\gamma_n=a=0.5\kappa$ and for inactive oscillators $\gamma_n=b=0.375\kappa$. The other parameters are $N=100, J=7.5\kappa$.}
	\label{fig6}
\end{figure}
Next, we investigate how the dissipative coupling influence the aging transition both in the classical and quantum regime. In the classical regime, the dependence of $p_\text{cmin}$ on the dissipative coupling strength $V$ can be directly obtained from Eq.~(\ref{equation6}). As shown in Fig.~\ref{fig6}(a), the critical ratio $p_c$ decreases monotonically as the dissipative coupling strength $V$ increase. Moreover, the aging transition can occur earlier (i.e., for smaller $p$) with respect to the case of $g=0$. To further characterize this behavior, we present the aging scenario in the $g-V$ parameter space by analyzing the variation of the normalized order parameter $Q_c$. As shown in Fig.~\ref{fig6}(b), at a fixed $p=0.6$, we observe a sharp boundary between the oscillatory state ($Q_c\neq0$) and the aging state ($Q_c=0$). By carefully tuning the atom-oscillator coupling $g$ and dissipative coupling $V$, the oscillators can be avoided to transition to the aging state, maintaining oscillations for the proper functioning of the system. In the quantum regime, Fig.~\ref{fig6}(c) shows that $p_c$ increases with increasing $V$, contrasting sharply with its classical counterpart where $p_c$ decreases with $V$. Then we present the quantum aging features  in the $g-V$ parameter space by visualizing the variation of the normalized order parameter $Q_q$. As shown in Fig.~\ref{fig6}(d), choosing $p=0.3$, we can observe that for a fixed $g$, there exists a specific value of $V$ that minimizes $Q_q$. Similarly, by appropriately selecting the atom-oscillator coupling strength $g$ and the dissipative coupling strength $V$, the order parameter can be controlled at a higher value.

\begin{figure}
	\centering
	\includegraphics[width=0.47\textwidth]{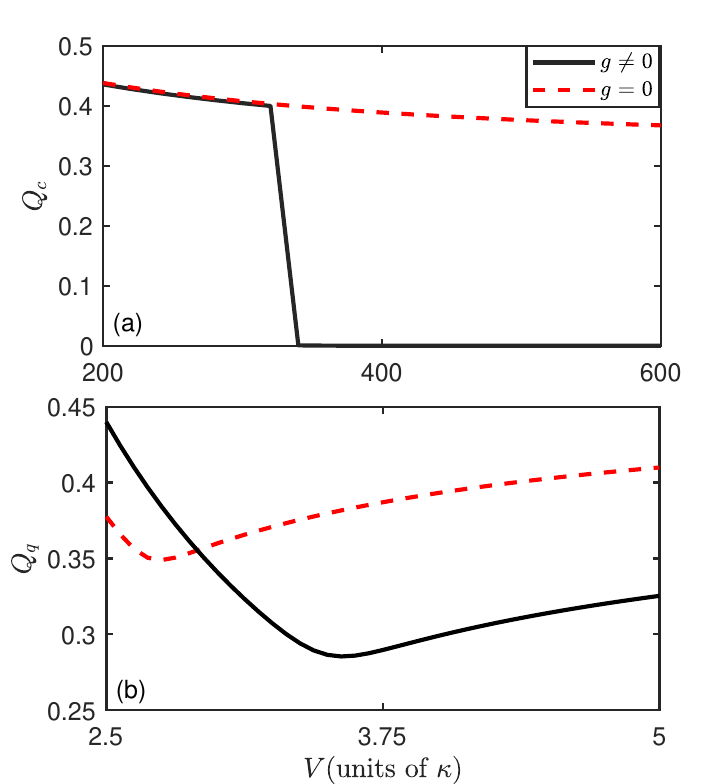}
	\caption {(a) Variation of order parameter $Q_c$ with $V$ in the presence and absence of atom-oscillator interaction at $p=0.6$. For active oscillators $\gamma_n=a=80\kappa$ and for inactive oscillators $\gamma_n=b=40\kappa$. The other parameters are $N=100, J=300\kappa, g=3.2\kappa$. (b) Variation of order parameter $Q_q$ with $V$ in the presence and absence of atom-oscillator interaction at $p=0.32$. For active oscillators $\gamma_n=a=0.5\kappa$ and for inactive oscillators $\gamma_n=b=0.375\kappa$. The other parameters chosen are $N=100, J=7.5\kappa, g=0.03\kappa$.}
	\label{fig7}
\end{figure}

Finally, we show the plot of $Q$ versus $V$ to explore how the results with atom-oscillator coupling differ from previously established results without it.  In the classical regime [as shown in Fig.~\ref{fig7}(a)], for $p=0.6$, the order parameter decreases monotonically with $V$ when there is no atom-oscillator coupling in the system. However, when $g\neq0$, it is evident that beyond a certain value of $V$, $Q_c$ undergoes a sudden drop to a globally stopped state. The sudden drop can be interpreted as a dependence of the system's dynamics on initial conditions: even a slight shift of $V$ can significantly alter the state to which the system ultimately converges. This phenomenon parallels the discontinuous transitions observed in the $Q_c-p$ plot. In the quantum regime [as shown in Fig.~\ref{fig7}(b)], specifically for $p=0.32$, we observe that $Q_q$ first decreases with $V$, reaches a minimum, and then increases when no atom-oscillator coupling is present ($g=0$). When $g\neq0$, this non-monotonic trend persists but shifts: the minimum $Q_q$ occurs at a larger $V$.

\renewcommand{\thesection}{\Roman{section}}
\section{CONCLUSION and discussions}\label{sec4}
In this manuscript, we investigated the aging transition in a network consisting of active and inactive oscillators coupled to a two-level atom. We have demonstrated that in both classical and quantum regime atom-oscillator couplings can modulate the aging transition point of the coupled oscillators. For classical oscillators, we have analytically derived the threshold of the ratio of the inactive oscillators to total oscillators for the aging transition to occur. Our results suggest that, with the parameters chosen, the threshold decreases with the atom-oscillator coupling strength increasing. This implies that in the presence of atom-oscillator coupling the aging transition can occur with a smaller number of inactive oscillators with respect  to the oscillators without atom-oscillator coupling. In addition, the atom-oscillator interaction cause the sudden suppression of the macroscopic oscillations in the system. We have shown that the similar effect could also be found in the quantum regime. The aging transition occurs at a lower value of $p$ when explicit atom-oscillator coupling is introduced, compared to the quantum oscillators without atom-oscillator coupling. The findings can be interpreted as the atom modulating the coupling strength of each oscillator to the mean field, which results in a reduced effective coupling strength $V$ compared to the case where $g=0$. Note that the critical value $p_c$ of the aging transition depends on the decay rate of the atom regardless of whether the oscillators are in classical or quantum regime.

The system can be physically realized with cavity optomechanical systems \cite{Walt2014,Walt2015} or trapped ions \cite{Lee2013}. The gain and loss  for active and inactive oscillators can be engineered using blue and red sidebands. Namely, the blue-detuned laser is set on the single-boson (e.g., photon, phonon, etc.) sideband, realizing one boson gain (i.e., active oscillator). The  red-detuned laser is set on the single-boson sideband and two-boson sideband, which induces one boson loss (inactive oscillator) and two boson absorption process (for both active and inactive oscillators). The global dissipative interaction between oscillators may be accomplished by positioning dielectric membranes in a Fabry-P$\acute{\text e}$rot cavity with a large out-coupling and by leveraging multiple modes of the cavity \cite{Walt2015}. Besides, the current model assumes uniform pairwise coupling between oscillators, but realistic complex systems often exhibit richer couplings. For instance, in our setup, nodes within the same subsystems can be highly interconnected, whereas connections to nodes in other type of subsystems  are sparse. One can consider the other topology such non-homogeneous structures in future works.

\section*{ACKNOWLEDGMENTS}\label{sec5}
This work is supported by National Natural Science Foundation of China (NSFC) under Grant No. 12175033.

\appendix

\section{COMPARISON OF THE RESULTS GIVEN BY THE CLASSICAL AND QUANTUM REGIME}\label{appendixA}
\begin{figure}[H]
	\centering
	\includegraphics[width=0.4\textwidth]{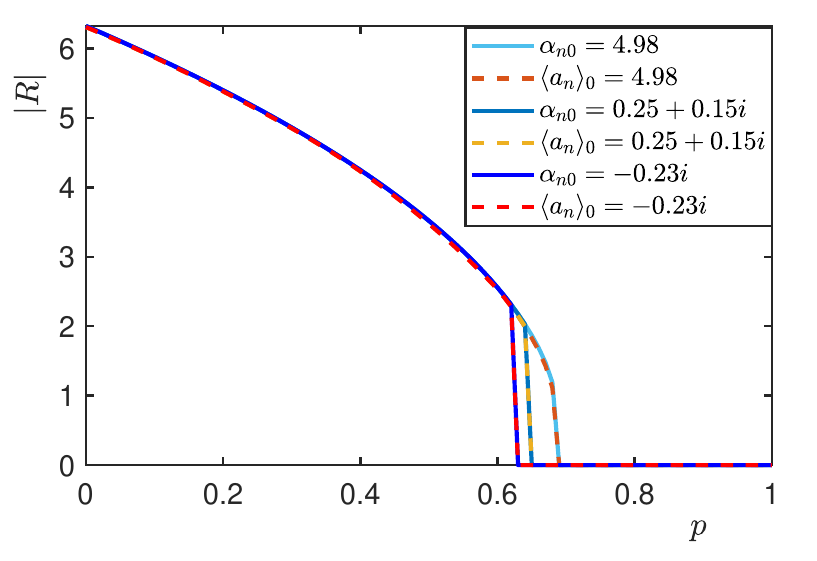}
	\caption {The order parameter $|R|$ as a function of the ratio of inactive to the total oscillators $p$ for different values of initial conditions ${\langle a_n\rangle}_0$ and $\alpha_{n0}$. The solid line represents the results in the classical regime, and the dashed line represents that in the quantum regime. For active oscillators $\gamma_n=a=80\kappa$ and for inactive oscillators $\gamma_n=b=40\kappa$. The other parameters chosen are $N=100, g=3.6\kappa, J=300\kappa, V=300\kappa$.}
	\label{fig8}
\end{figure}
In this Appendix, we will demonstrate that the approximations made in the main text by considered the quantum oscillators as classical when $\kappa\ll\gamma_n$ is a good approximation. In the classical regime, the expectation $\langle a_n\rangle$ can be replaced with a complex amplitude  $\alpha_n$.  We compare the results obtained through numerical simulations based on Eq.~(\ref{equation3}) and Eq.~(\ref{equation7}). Initially, the atom is in the excited state and all oscillators are set to the same coherent state $|\alpha_0\rangle$. Therefore, the initial conditions for oscillators both in quantum and classical regime are ${\langle a_n\rangle}_0=\alpha_{n0}=\alpha_0$. We have randomly generated several initial conditions $\alpha_0$ within the range of $|\alpha_0|=0\thicksim5$ and plotted the order parameter $|R|$ versus $p$ in the classical and quantum domains, respectively. In the quantum regime, $R\equiv N^{-1}\sum_{n=1}^{N}\langle a_n\rangle;$ in the classical regime, $R\equiv N^{-1}\sum_{n=1}^{N}\alpha_n$. From Fig.~\ref{fig8}, it can be observed that the results based on the classical equation (\ref{equation3}) and quantum equation (\ref{equation7}) are almost agreement under different initial conditions. This confirms the validity of our approximation.

\section{THE REASON FOR THE SAME DECAY RATE OF $Q_c$ FOR DIFFERENT VALUES OF THE ATOM-OSCILLATOR COUPLING STRENGTH}\label{appendixB}
In the classical regime, we can replace the expectation $\langle a_n\rangle$ with a complex value $\alpha_n$, as well as $\langle a_{n}^\dag a_{n}^2\rangle\rightarrow|\alpha_n|^2\alpha_n$ in the equation of motion for $\langle a_n\rangle$. Assuming that in each group, all oscillators are in the same state and setting $\alpha_n=A$ and $\gamma_n=a$ for all active oscillators as well as $\alpha_n=I$ and $\gamma_n=b$  for all inactive oscillators, we obtain
\begin{equation}
\begin{aligned}
\dot{\langle\sigma_{-}\rangle}&=-ig\sum_{n=1}^{N}\alpha_n
(1-2\langle\sigma_{+}\sigma_{-}\rangle)-\frac{J}{2}\langle\sigma_{-}\rangle,\\
\dot{\langle\sigma_{+}\sigma_{-}\rangle}&=-ig\sum_{n=1}^{N}(\alpha_n\langle\sigma_{+}\rangle-\alpha_{n}^{*}
\langle\sigma_{-}\rangle)-J\langle\sigma_{+}\sigma_{-}\rangle,\\
\dot{A}&=-ig\langle\sigma_{-}\rangle+(\frac{a}{2}-Vp-\kappa|A|^2)A+VpI,\\
\dot{I}&=-ig\langle\sigma_{-}\rangle+(-\frac{b}{2}-Vq-\kappa|I|^2)I+VqA,
\label{reply_eq1}
\end{aligned}
\end{equation}
where $q\equiv1-p$. We analyze the steady-state by solving $\dot{\langle\sigma_{-}\rangle} = \dot{\langle\sigma_{+}\sigma_{-}\rangle} = \dot{A} = \dot{I} = 0$, which leads to a set of equations for $A$ and $I$
\begin{equation}
\begin{aligned}
(\frac{a}{2}-Vp-\kappa|A|^2)A+VpI-\frac{2g^2J/Z^*}{J^2/|Z|^2+8g^2}&=0,\\
(-\frac{b}{2}-Vq-\kappa|I|^2)I+VqA-\frac{2g^2J/Z^*}{J^2/|Z|^2+8g^2}&=0,
\label{reply_eq2}
\end{aligned}
\end{equation}
$Z\equiv\sum_{n=1}^{N}\alpha_n=N(1-p)A+NpI$, which represents the sum of the amplitudes of $N$ oscillators. With the parameters range chosen, $Z$ is large enough before $p$ reaches the threshold $p_\text{cmin}$, causing $\frac{J^2}{|Z|^2}\ll8g^2$. Therefore, Eq.~(\ref{reply_eq2}) can be further approximated as
\begin{equation}
\begin{aligned}
(\frac{a}{2}-Vp-\kappa|A|^2)A+VpI-\frac{J}{4Z^*}&=0,\\
(-\frac{b}{2}-Vq-\kappa|I|^2)I+VqA-\frac{J}{4Z^*}&=0,
\label{reply_eq3}
\end{aligned}
\end{equation}
From Eq.~(\ref{reply_eq3}), we find that the magnitude of $A$ and $I$ does not depend on $g$, leading to the same decay rate of the normalized order parameter $Q_c$ for different values of the atom-oscillator coupling strength $g$. We also verify that the results obtained by using Eq.~(\ref{reply_eq3}) are basically consistent with Eq.~(\ref{reply_eq1}) before $p$ reaches the threshold $p_\text{cmin}$. We need to emphasize that the aging transition can occur only if the inactive ratio exceeds the threshold, and the threshold $p_\text{cmin}$   needs to be determined by stability analysis and it depends on $g$.

\section{THE FOCK DISTRIBUTION OF THE QUANTUM ACTIVE AND INACTIVE OSCILLATORS AFTER THE SYSTEM UNDERGOES AGING TRANSITION}\label{appendixC}
Now we write out Eq.~(\ref{equation7}) in terms of matrix elements. In the quantum regime, we consider $\kappa>\gamma_n$ and truncate  the Hilbert space with boson number at most $n_0=4$. Using the notation $\rho_{j,k}\equiv\langle j|\rho|k\rangle$, the equations of motion for the density matrix elements are given by
\begin{widetext}
\begin{eqnarray}
\label{A1}
\dot{\rho_{e,g}^a}&=&-ig\sum_{n=1}^{N}{\langle a\rangle}_n(1-2\rho_{e,e}^a)-\frac{J}{2}\rho_{e,g}^a,\\
\dot{\rho_{e,e}^a}&=&-ig\sum_{n=1}^{N}({\langle a\rangle}_n\rho_{g,e}^a-{\langle a^\dagger\rangle}_n\rho_{e,g}^a)-J\rho_{e,e}^a,\\
\dot{\rho_{j,k}^A}&=&\frac{\gamma_n}{2}[2\sqrt{jk}\rho_{j-1,k-1}^A-(j+k+2)\rho_{j,k}^A]
+\frac{V(N-1)}{N}[2\sqrt{(j+1)(k+1)}\rho_{j+1,k+1}^A-(j+k)\rho_{j,k}^A]\nonumber\\
&+&[\frac{V}{N}M_A-ig\rho_{e,g}^a](\sqrt{j}\rho_{j-1,k}^A-\sqrt{k+1}\rho_{j,k+1}^A)-[\frac{V}{N}M_A^{*}+ig\rho_{g,e}^a]
(\sqrt{j+1}\rho_{j+1,k}^A-\sqrt{k}\rho_{j,k-1}^A)\nonumber\\
&+&\frac{\kappa}{2}[2\sqrt{(j+1)(j+2)(k+1)(k+2)}\rho_{j+2,k+2}^A-[j(j-1)+k(k-1)]\rho_{j,k}^A],\\
\dot{\rho_{j,k}^I}&=&\frac{\gamma_n}{2}[2\sqrt{(j+1)(k+1)}\rho_{j+1,k+1}^I-(j+k)\rho_{j,k}^I]
+\frac{V(N-1)}{N}[2\sqrt{(j+1)(k+1)}\rho_{j+1,k+1}^I-(j+k)\rho_{j,k}^I]\nonumber\\
&+&[\frac{V}{N}M_I-ig\rho_{e,g}^a](\sqrt{j}\rho_{j-1,k}^I-\sqrt{k+1}\rho_{j,k+1}^I)-[\frac{V}{N}M_I^{*}+ig\rho_{g,e}^a]
(\sqrt{j+1}\rho_{j+1,k}^I-\sqrt{k}\rho_{j,k-1}^I)\nonumber\\
&+&\frac{\kappa}{2}[2\sqrt{(j+1)(j+2)(k+1)(k+2)}\rho_{j+2,k+2}^I-[j(j-1)+k(k-1)]\rho_{j,k}^I],
\end{eqnarray}
\end{widetext}
For $n_0=4$, the mean fields can be written as
\begin{equation}
\begin{aligned}
\sum_{n=1}^{N}{\langle a\rangle}_n&=N(1-p)(\rho_{10}^A+\sqrt{2}\rho_{21}^A+\sqrt{3}\rho_{32}^A+2\rho_{43}^A)\\&+
Np(\rho_{10}^I+\sqrt{2}\rho_{21}^I+\sqrt{3}\rho_{32}^I+2\rho_{43}^I),\\
M_A&=[N(1-p)-1](\rho_{10}^A+\sqrt{2}\rho_{21}^A+\sqrt{3}\rho_{32}^A+2\rho_{43}^A)\\&+
Np(\rho_{10}^I+\sqrt{2}\rho_{21}^I+\sqrt{3}\rho_{32}^I+2\rho_{43}^I),\\
M_I&=N(1-p)(\rho_{10}^A+\sqrt{2}\rho_{21}^A+\sqrt{3}\rho_{32}^A+2\rho_{43}^A)\\&+
(Np-1)(\rho_{10}^I+\sqrt{2}\rho_{21}^I+\sqrt{3}\rho_{32}^I+2\rho_{43}^I),
\label{A2}
\end{aligned}
\end{equation}
We now find the steady-solution of the aging state, denoted by $\bar{\rho}$, which is a fixed point of the mean-field equations [Eq.~(\ref{equation7})] but with no off-diagonal elements. The mean-field equations change from nonlinear to linear. From Eq.~(\ref{A1})-(\ref{A2}), we can easily find that $\overline{\rho_{00}^I}=1$, and the population of the inactive oscillator is 0 in all other states.  The Fock distribution of active oscillators is easily found by solving  the equations of motion for the diagonal elements $\dot{\rho_{j,j}^A}=0$
\begin{equation}
\begin{aligned}
&-\gamma_n\rho_{00}^A+2\kappa\rho_{22}^A+\frac{2V(N-1)}{N}\rho_{11}^A=0,\\
&\gamma_n(\rho_{00}^A-2\rho_{11}^A)+6\kappa\rho_{33}^A+\frac{2V(N-1)}{N}(2\rho_{22}^A-\rho_{11}^A)=0,
\nonumber
\end{aligned}
\end{equation}
\begin{equation}
\begin{aligned}
&\gamma_n(2\rho_{11}^A-3\rho_{22}^A)+2\kappa(6\rho_{44}^A-\rho_{22}^A)\\&+\frac{2V(N-1)}{N}(3\rho_{33}^A-2\rho_{22}^A)=0,\\
&\gamma_n(3\rho_{22}^A-4\rho_{33}^A)-6\kappa\rho_{33}^A+\frac{2V(N-1)}{N}(4\rho_{44}^A-3\rho_{33}^A)=0,\\
&4\gamma_n\rho_{33}^A-12\kappa\rho_{44}^A-\frac{8V(N-1)}{N}\rho_{44}^A=0,
\end{aligned}
\end{equation}
Through simple calculations, we can obtain
\begin{equation}
\begin{aligned}
\overline{\rho_{00}^A}&=[4\gamma_n\kappa^2(2\gamma_n+3\kappa)c_1+2\kappa^2(23\gamma_n+6\kappa)c_2\\&
+4\kappa(9\gamma_n+11\kappa)c_3+48\kappa c_4+16c_5]/C,\\
\overline{\rho_{11}^A}&=[\gamma_n\kappa(6\kappa^2+13\gamma_n\kappa)c_1+\gamma_n\kappa(22\kappa+14\gamma_n)c_2\\&
+24\gamma_n\kappa c_3+8\gamma_n c_4]/C,\\
\overline{\rho_{22}^A}&=[2\gamma_n\kappa(2{\gamma_n}^2+3\gamma_n\kappa)c_1+10{\gamma_n}^2\kappa c_2+4{\gamma_n}^2 c_3]/C,\\
\overline{\rho_{33}^A}&=(3{\gamma_n}^3\kappa c_1+6{\gamma_n}^3 c_2)/C,\\
\overline{\rho_{44}^A}&={\gamma_n}^4c_1/C,
\label{A3}
\end{aligned}
\end{equation}
where $c_1=N^4, c_2=(N-1)N^3V, c_3=(N-1)^2N^2V^2, c_4=(N-1)^3NV^3, c_5=(N-1)^4V^4$. $C=\gamma_n({\gamma_n}^3+7{\gamma_n}^2\kappa+27{\gamma_n}\kappa^2+18\kappa^3)c_1+2({\gamma_n}^3+12{\gamma_n}^2\kappa+
34\gamma_n\kappa^2+6\kappa^3)c_2+4({\gamma_n}^2+15\gamma_n\kappa+11\kappa^2)c_3+8(\gamma_n+6\kappa)c_4+16c_5$. All off-diagonal elements are zero, since this implies the lack of coherence among the oscillators. The analytical expressions for the Fock distribution of the active  oscillators are given by Eq.~(\ref{A3}).


\vspace*{0.5cm}
\begin{thebibliography}{10}

\bibitem{Ge2004}H. X. Ge, S. Q. Dai, L. Y. Dong, and Y. Xue, Stabilization effect of traffic flow in an extended car-following model based on an intelligent transportation system application, Phys. Rev. E \textbf{70}, 066134 (2004).

\bibitem{Rohden2012}M. Rohden, A. Sorge, M. Timme, and D. Witthaut, Self-Organized Synchronization in Decentralized Power Grids, Phys. Rev. Lett. \textbf{109}, 064101 (2012).

\bibitem{Cama2003}S. Camazine, \textit{Self-organization in Biological Systems} (Princeton University Press, Princeton, NJ, 2003).

\bibitem{Hopf1982}J. J. Hopfield, Neural networks and physical systems with emergent collective computational abilities, Proc. Natl. Acad. Sci. USA \textbf{79}, 2554 (1982).

\bibitem{Piko2001}A. S. Pikovsky, M. Rosenblum, and J. Kurths, \textit{Synchronization: A Universal Concept in Nonlinear Science} (Cambridge University Press, New York, 2001).

\bibitem{Xu2019}C. Xu, J. Gao, S. Boccaletti, Z. Zheng, and S. Guan, Synchronization in starlike networks of phase oscillators, Phys. Rev. E \textbf{100}, 012212 (2019).

\bibitem{Xie2014}J. Xie, E. Knobloch, and H.-C. Kao, Multicluster and traveling chimera states in nonlocal phase-coupled oscillators, Phys. Rev. E \textbf{90}, 022919 (2014).

\bibitem{Kori2014}H. Kori, Y. Kuramoto, S. Jain, I. Z. Kiss, and J. L. Hudson, Clustering in globally coupled oscillators near a Hopf bifurcation: Theory and experiments, Phys. Rev. E \textbf{89}, 062906 (2014).

\bibitem{Jaros2018}P. Jaros, S. Brezetsky, R. Levchenko, D. Dudkowski, T. Kapitaniak, and Y. Maistrenko, Solitary states for coupled oscillators with inertia, Chaos \textbf{28}, 011103 (2018).

\bibitem{Teich2019}E. Teichmann and M. Rosenblum, Solitary states and partial synchrony in oscillatory ensembles with attractive and repulsive interactions, Chaos \textbf{29}, 093124 (2019).

\bibitem{Sath2018}K. Sathiyadevi, V. K. Chandrasekar, D. V. Senthilkumar, and M. Lakshmanan, Distinct collective states due to trade-off between attractive and repulsive couplings, Phys. Rev. E \textbf{97}, 032207 (2018).

\bibitem{Sathi2019}K. Sathiyadevi, V. K. Chandrasekar, D. V. Senthilkumar, and M. Lakshmanan, Long-range interaction induced collective dynamical behaviors, J. Phys. A: Math. Theor. \textbf{52}, 184001 (2019).

\bibitem{Kura1996}Y. Kuramoto and H. Nakao, Origin of power-law spatial correlations in distributed oscillators and maps with nonlocal coupling, Phys. Rev. Lett. \textbf{76}, 4352 (1996).

\bibitem{Kura1997}Y. Kuramoto and H. Nakao, Power-law spatial correlations and the onset of individual motions in self-oscillatory media with non-local coupling, Physica D \textbf{103}, 294 (1997).

\bibitem{Zhu2014}Y. Zhu, Z. Zheng, and J. Yang. Chimera states on complex networks, Phys. Rev. E \textbf{89}, 022914 (2014).
\bibitem{Gopal2014}R. Gopal, V. K. Chandrasekar, A. Venkatesan, and M. Lakshmanan, Observation and characterization of chimera states in coupled dynamical systems with nonlocal coupling, Phys. Rev. E \textbf{89}, 052914 (2014).

\bibitem{Sax2012}G. Saxena, A. Prasad, and R. Ramaswamy, Amplitude death: The emergence of stationarity in coupled nonlinear systems, Phys. Rep. \textbf{521}, 205 (2012).

\bibitem{Schn2015}I. Schneider, M. Kapeller, S. Loos, A. Zakharova, B. Fiedler, and E. Sch$\ddot{\text{o}}$ll, Stable and transient multicluster oscillation death in nonlocally coupled networks. Phys. Rev. E \textbf{92}, 052915 (2015).

\bibitem{Baner2018}T. Banerjee, D Biswas, D Ghosh, B Bandyopadhyay, and J Kurths, Transition from homogeneous to inhomogeneous limit cycles: effect of local filtering in coupled oscillators, Phys. Rev. E \textbf{97}, 042218 (2018).

\bibitem{Daido2004}H. Daido and K. Nakanishi, Aging Transition and Universal Scaling in Oscillator Networks, Phys. Rev. Lett. \textbf{93}, 104101 (2004).

\bibitem{Bandy2023}B. Bandyopadhyay and T. Banerjee, Aging transition in coupled quantum oscillators, Phys. Rev. E \textbf{107}, 024204 (2023).

\bibitem{Zhang2024}H. Zhang, D. Cui, W. Wang, and X. X. Yi, Aging of coupled qubits, Phys. Rev. A \textbf{110}, 012221 (2024).

\bibitem{Majhi2024}S. Majhi, B. Rakshit, A. Sharma, J. Kurths, and D. Ghosh, Dynamical robustness of network of oscillators, Phys. Rep. \textbf{1082}, 1 (2024).

\bibitem{Daido2007}H. Daido and K. Nakanishi, Aging and clustering in globally coupled oscillators, Phys. Rev. E \textbf{75}, 056206 (2007).

\bibitem{Daido2008}H. Daido, Aging transition and disorder-induced coherence in locally coupled oscillators, EPL \textbf{84}, 10002 (2008).

\bibitem{Sun2019}Z. Sun, Y. Liu, K. Liu, X. Yang, and W. Xu, Aging transition in mixed active and inactive fractional-order oscillators, Chaos \textbf{29}, 103150 (2019).

\bibitem{Sun2017}Z. Sun, N. Ma, and W. Xu, Aging transition by random errors, Sci. Rep. \textbf{7}, 42715 (2017).

\bibitem{Sath2019}K. Sathiyadevi, I. Gowthaman, D. V. Senthilkumar and V. K. Chandrasekar, Aging transition in the absence of inactive oscillators, Chaos \textbf{29}, 123117 (2019).

\bibitem{Sath2022}K. Sathiyadevi, D. Premraj, T. Banerjee, Z. Zheng, and M. Lakshmanan, Aging transition under discrete time-dependent coupling: Restoring rhythmicity from aging, Chaos, Solitons Fractals \textbf{157}, 111944 (2022).

\bibitem{Ponrasu2020}K. Ponrasu, I. Gowthaman, V. K. Chandrasekar, and D. V. Senthilkumar, Aging transition under weighted conjugate coupling, EPL \textbf{128}, 58003 (2019).

\bibitem{Biswas2022}D. Biswas and S. Gupta, Ageing transitions in a network of Rulkov neurons, Sci. Rep. \textbf{12}, 433 (2022).

\bibitem{Sahoo2023}S. Sahoo, A. Prasad and R. Ramaswamy, Stasis in heterogeneous networks of coupled oscillators:
discontinuous transition with hysteresis, J. Phys. Complex. \textbf{4}, 035001 (2023).

\bibitem{Singh2020}U. Singh, K. Sathiyadevi, V. K. Chandrasekar, W. Zou, J. Kurths, and D. V. Senthilkumar, Trade-off between filtering and symmetry breaking mean-field coupling in inducing macroscopic dynamical states, New J. Phys. \textbf{22} 093024 (2020).

\bibitem{Rahman2017}B. Rahman, K. B. Blyuss, and Y. N. Kyrychko, Aging transition in systems of oscillators with global distributed-delay coupling, Phys. Rev. E \textbf{96}, 032203 (2017).

\bibitem{Rakshit2020}B. Rakshit, N. Rajendrakumar, and B. Balaram, Abnormal route to aging transition in a network
of coupled oscillators, Chaos \textbf{30}, 101101 (2020).

\bibitem{Zou2021}W. Zou, D. V. Senthilkumar, M. Zhan, and J. Kurths, Quenching, aging, and reviving in coupled dynamical networks, Phys. Rep. \textbf{931}, 1 (2021).

\bibitem{Hei2014}T. T. Heikkil$\ddot{\text{a}}$, F. Massel, J. Tuorila, R. Khan, and M. A. Sillanp$\ddot{\text{a}}\ddot{\text{a}}$, Enhancing Optomechanical Coupling via the Josephson Effect, Phys. Rev. Lett. \textbf{112}, 203603 (2014).

\bibitem{Pirk2015}J.-M. Pirkkalainen, S. U. Cho, F. Massel, J. Tuorila, T. T. Heikkil$\ddot{\text{a}}$, P. J. Hakonen, and M. A. Sillanp$\ddot{\text{a}}\ddot{\text{a}}$, Cavity Optomechanics Mediated by a Quantum Two-Level System, Nat. Commun. \textbf{6}, 6981 (2015).

\bibitem{Liao2014}J.-Q. Liao, K. Jacobs, F. Nori, and R. W. Simmonds, Modulated Electromechanics: Large Enhancements of Nonlinearities, New J. Phys. \textbf{16}, 072001 (2014).

\bibitem{Li2020}P.-B. Li, Y. Zhou, W.-B. Gao, and F. Nori, Enhancing spin-phonon and spin-spin interactions using linear resources in a hybrid quantum system, Phys. Rev. Lett. \textbf{125}, 153602 (2020).

\bibitem{Duan2004}L. M. Duan and H. J. Kimble, Scalable photonic quantum computation through cavity-assisted interactions, Phys. Rev. Lett. \textbf{92}, 127902 (2004).

\bibitem{Duan2005}L. M. Duan, B. Wang, and H. J. Kimble, Robust quantum gates on neutral atoms with cavity-assisted photon scattering, Phys. Rev. A \textbf{72}, 032333 (2005).

\bibitem{Hwang2015}M.-J. Hwang, R. Puebla, and M. B. Plenio, Quantum Phase Transition and Universal Dynamics in the Rabi Model, Phys. Rev. Lett. \textbf{115}, 180404 (2015).

\bibitem{Lee2014}T. E. Lee, C.-K. Chan, and S. Wang, Entanglement tongue and quantum synchronization of disordered oscillators, Phys. Rev. E \textbf{89}, 022913 (2014).

\bibitem{Ishi2017}K. Ishibashi and R. Kanamoto, Oscillation collapse in coupled quantum van der Pol oscillators, Phys. Rev. E \textbf{96}, 052210 (2017).

\bibitem{Kubo1962}R. Kubo, Generalized cumulant expansion method, J. Phys. Soc. Jpn. \textbf{17}, 1100 (1962).

\bibitem{Strogatz1994}S. H. Strogatz, \textit{Nonlinear Dynamics and Chaos} (Perseus Books, Cambridge, 1994).

\bibitem{Liu1994}W.-M. Liu, Criterion of Hopf bifurcations without using eigenvalues, J. Math. Anal. Appl. \textbf{182}, 250 (1994).

\bibitem{Lugiato2015}L. Lugiato, F. Prati, and M. Brambilla, \textit{Nonlinear Optical Systems} (Cambridge University Press, 2015).

\bibitem{Lee2013}T. E. Lee and H. R. Sadeghpour, Phys. Rev. Lett. \textbf{111}, 234101 (2013).

\bibitem{Walt2014}S. Walter, A. Nunnenkamp, and C. Bruder, Phys. Rev. Lett. \textbf{112}, 094102 (2014).

\bibitem{Walt2015}S. Walter, A. Nunnenkamp, and C. Bruder, Ann. Phys. (Berlin) \textbf{527}, 131 (2015).


\end{thebibliography}
\end{document}